\documentclass[english,prl]{revtex4-1}
\usepackage[T1]{fontenc}
\usepackage[latin9]{inputenc}
\usepackage{graphicx}

\makeatletter

\@ifundefined{textcolor}{}
{%
 \definecolor{BLACK}{gray}{0}
 \definecolor{WHITE}{gray}{1}
 \definecolor{RED}{rgb}{1,0,0}
 \definecolor{GREEN}{rgb}{0,1,0}
 \definecolor{BLUE}{rgb}{0,0,1}
 \definecolor{CYAN}{cmyk}{1,0,0,0}
 \definecolor{MAGENTA}{cmyk}{0,1,0,0}
 \definecolor{YELLOW}{cmyk}{0,0,1,0}
}

\makeatother

\usepackage{babel}

\begin{document}
\vspace*{0.2in}

\begin{flushleft}
{\Large
\textbf\newline{Frequency Dependent Visco-elastic Properties of a Water Nanomeniscus: an AFM Study in Force Feedback Mode} 
}
\newline
\\
Simon Carpentier\textsuperscript{1,2},
Mario S Rodrigues\textsuperscript{3},
Luca Costa\textsuperscript{4},
Miguel V Vitorino\textsuperscript{3},
Jo\"{e}l Chevrier\textsuperscript{1,2}
\\
\bigskip
\textbf{1} Univ. Grenoble Alpes, F-38000 Grenoble, France
\\
\textbf{2} CNRS, Inst NEEL, F-38042 Grenoble, France
\\
\textbf{3} University of Lisboa, Falculty of Sciences, BioISI-Biosystems \& Integrative Sciences Institute, Campo Grande, Lisboa, P-1749-016, Portugal
\\
\textbf{4} ESRF, The European Synchrotron, 71 Rue des Martyrs, 38000 Grenoble, France
\\
\bigskip

%
%





*simon.carpentier@neel.cnrs.fr

\end{flushleft}
\section*{Abstract}
Recently, using an Atomic Force Microscope and a single cantilever
excited at different frequencies it was shown that water nanomeniscus
can exhibit a transition in mechanical responses when submitted to
stimuli above few tens of kHz. The use of a single cantilever to explore
phenomena at frequencies far from the cantilever resonance frequency
is not a common and well-established strategy, and because water meniscus
are ubiquitous in nature, we have also studied the water meniscus
mechanical response, the stiffness $G'$ (N/m) and
the dissipation $G''$ (kg/s), as a function of frequency
by using cantilevers with different resonance frequencies. These results,
based on classic dynamical AFM technics, confirm the anomalous mechanical
response of water nanomeniscus when stimulated at frequencies high
enough.


\section*{Introduction}

Capillary condensation occurs as soon as a confined geometry is in a presence of a vapor 
phase~\cite{thomson18724,barcons2012nanoscale,Szoszkiewicz2005,riedo2002kinetics,mugele2002capillarity}. 
Capillary forces are among the most intense at nanoscale, 
playing an important role in the interaction between almost all surfaces.
The importance of dynamical properties of water bridges becomes evident 
if one considers interacting surfaces with roughness scales down to few nanometers, since in that case,
even at moderate speeds of m/s sliding will generate frequencies above 1 MHz. 
Between an Atomic Force Microscope (AFM) tip and a surface, this interaction, in a long time range, 
is attractive and remains at the thermodynamical equilibrium. 
It is described by the Young-Laplace equation, 
The nanomeniscus has to adapt its radii of curvature to keep the chemical potentials equal. 
We have highlighted that a nanomeniscus can appear as a stiff material (0.1GPa) 
at the nanoscale when excited at high frequency~\cite{carpentier2015out}. 
This was observed using an AFM mode called Force Feedback Microscopy (FFM)~\cite{rodrigues2012atomic}. 
We have argued that molecular exchange progressively loses efficiency, 
when the frequency of the tip oscillation is increased and thus the nanomeniscus is no longer at the thermodynamical equilibrium. 
The capillary bridge is then led to acquire a constant volume during oscillations at high frequencies.
In addition to the constant volume picture, the ability of the bridge to adapt to a new shape looses efficiency
resulting in a progressive locking of the contact line resulting in changes of the mean curvature radius of the capillary bridge
This appears as a transition between a negative stiffness $G'$ (N/m) in the long time regime to positive stiffness at high frequency. 
The estimated characteristic time is $10^{-6}$s. 

Our work published in Ref~\cite{carpentier2015out} was a first approach 
to this question based on a non-conventional use of AFM. 
It is important to stress that even though the methodology used 
before is sound it is not well established among the AFM community.
Here we want to show that the key results reported in our previous work, 
about dynamical properties of water nanobridge, 
can be, though only partially, obtained using well-known AFM strategies such as Amplitude Modulation AFM (AM-AFM).
These classical AFM measurements must, however, be combined with Force Feedback Mode to ensure tip stability. 
Because we have used a tip oscillation with amplitude smaller or equal to 1nm, a linear analysis of measurements 
of amplitude and phase can be done to to extract the dynamical properties of the water nanobridge. 
Compared to our work in Ref~\cite{carpentier2015out}, what we report here is a limited approach as it requires comparison 
of data from different experiments and obtained using different probes. 
The results here reported confirm previous measurements and further ascertain the associated conclusions. 

We have probed dynamical properties of water nanomeniscus at two different frequencies 79kHz
and 350kHz using two different AFM levers. 
During all these experiments, a force feedback ensures stability of the tip position. 
This leads to a direct determination of the static force applied by the nanobridge onto the tip,
and prevents jump to contact (i.e. there is no movement of the tip when the water nanomeniscus suddenly forms). 
A small oscillation is added at a frequency near the resonance frequency, 
which enables us to measure the phase and amplitude as a function of distance. 
Through a linear transformation of amplitude and phase, the nano-meniscus stiffness $G'$ and the associated dissipation $G''$
are obtained. 
With this method, we obtain a full characterization of the mediated interaction by the water nanobridge at these two frequencies.

\section*{Materials and Methods}

\begin{figure}[ht]
\includegraphics[width=15cm]{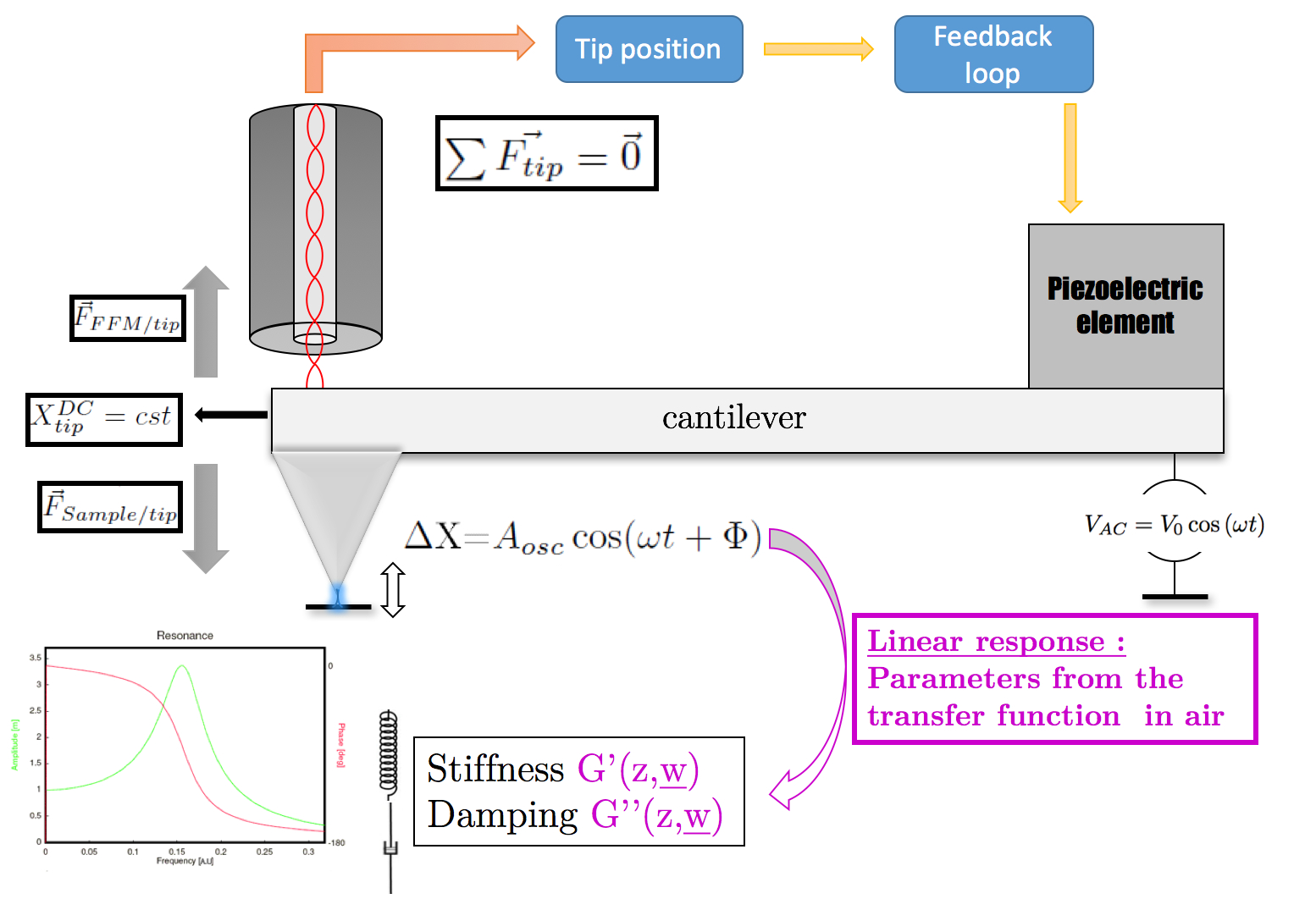}
\caption{Description of the FFM measurement strategy. 
The DC tip position is kept constant. 
This is due to real time application of a feedback force
determined through a control loop after measurement of the tip position.
An electrostatic coupling between the tip and the sample enforces
an AC tip oscillation with 1 nm of free oscillation amplitude. The
amplitude of oscillation $A_{osc}$ and the phase
$\Phi$ are the measured quantities.}
\label{fig:FFM_strategy}
\end{figure}

Throughout all presented measurements, the static force applied by the capillary
bridge to the tip is canceled in real time by a feedback force.
To apply this force in real time, a piezoelement changes the
DC position of the clamped end part of the microlever. The displacement
needed to maintain the tip position constant multiplied by the stiffness $k$
gives access to the static capillary force acting on the tip, for details on
experimental setup see Refs~\cite{rodrigues2014system,rodrigues2012atomic,carpentier2015proximity,vitorino2014force,costa2013comparison}.
The tip oscillation is detected by an optical fiber positioned close
to the cantilever backside, as depicted in Fig~\ref{fig:FFM_strategy}.

The sample is a flat piece of silicon wafer with its native oxide.
Two different levers were used, both with silicon tips:
cantilever 1 (Multi75Al) with a nominal stiffness
of 2.7 N/m, a resonance frequency of 79kHz and a quality factor in air of 127 and cantilever 2 (Tap300Al) with
a nominal stiffness of 40 N/m, a resonance frequency of 349kHz and
a quality factor of 200. Both cantilevers were purchased from BudgetSensors.

The experiment is performed at room conditions. The relative humidity was 40 \%. 
All measurements are done during tip retraction in presence of a water
nanobridge until it ruptures. 
The mean curvature radius of the bridge can be estimated by the distance at which the bridge forms. 
In these measurements it is typically 10nm.

The capillary bridge stiffness and dissipation 
can be obtained through the measurement of the resonance
frequency $f_0$ and the $Q$ factor. 
Since, at each distance, both the amplitude and phase can be measured as a function
of frequency, fitting the amplitude and phase curves with an harmonic oscillator model, 
results in two sets of the pair \{$f_0(z)$, $Q(z)$\}. 
Each set is obtained using Eq~\ref{eq:2 H(f)} and Eq~\ref{eq:1 phy} respectively. 

\begin{equation}
H(z)^2=\frac{A_0^2}{\left(1-\frac{f^2}{f_0^2(z)}\right)^2+\left(\frac{f}{Q(z)\,f_0(z)}\right)^2} \label{eq:2 H(f)}
\end{equation}

\begin{equation}
\Phi(z)=\arctan\left(\frac{f/f_0(z)}{Q(z) \left(1-f^2/f_0(z)^2 \right)} \right) \label{eq:1 phy}
\end{equation}

From each of these sets, it is possible to extract the interaction stiffness $G'$ and damping $G''$.
The capillary bridge dynamics is separated from the cantilever dynamics trough Eq~\ref{eq:3} and Eq~\ref{eq:4}.

\begin{equation}
G'(z)=k\left(\frac{f_0(z)^2}{f_0^{\infty 2}}-1 \right) \label{eq:3}
\end{equation}

\begin{equation}
G''(z)=\frac{k \left(Q^\infty f_0(z)-Q(z) f_0^\infty \right)}{ 2 \pi Q^{\infty} Q(z) f_0^{\infty 2}} \label{eq:4}
\end{equation}

The amplitude and phase are recorded using a lock-in amplifier, for each z-position for cantilever
1 (see Figure.\ref{fig:1}) and for cantilever 2 (see Figure.\ref{fig:2}).
In this measurement scheme, the system transfer function takes time
to record because the excitation frequency must be sweep  over a given frequency range, which
takes about one minute. On this time scale, it may have a significant thermal
drift of up to 2 nm which results in a slightly distorted z-axis. 
The knowledge of the static force curve (Fig.\ref{fig:1}a) gives access to the tip-sample distance
as the mean tip position is kept constant during all of the experiment.

$G'$ and $G''$ can be obtained a third time by setting the frequency constant, and then measuring the changes 
of phase and amplitude as a function of distance. 
This is equivalent to a force approach curve in amplitude modulation
mode~\cite{hansma1994tapping,san2001tip,erlandsson1998force}. 
In this linear regime, $G'$ and $G''$ are separated from lever dynamics by using Eq~\ref{eq:5} and Eq~\ref{eq:6} 
(details in Ref~\cite{rodrigues2014system}):

\begin{equation}
G'(z) = F_r \left[ n(z) \cos(\Phi(z))- \cos(\Phi_\infty) \right] \label{eq:5}
\end{equation}

\begin{equation}
G''(z) = \frac{F_r}{\omega_0^\infty} \left[ -n(z) \sin(\Phi(z))+\sin(\Phi_\infty) \right] \label{eq:6}
\end{equation}

Where $n$ is the normalized amplitude of oscillation, it is equal to 1 far from the sample and $\Phi$ is the phase. 
These two quantities are recorded at the same frequency during the whole
experiment, in this case this is the free resonance frequency $f_0^\infty$.
Parameters $F_r$ and $\Phi_\infty$ are obtained using the lever properties in the experimental environment
(see more details in Ref.\cite{rodrigues2014system}):

\begin{equation}
F_{r}=[(k-m\omega^{2})^{2}+\gamma^{2}\omega^{2}]^{1/2}\label{eq:Fr}
\end{equation}

\begin{equation}
\Phi_\infty = \arctan \left( \frac{\gamma \omega}{k- m \omega^2} \right)   \label{eq:phy infiny}
\end{equation}

Using the transfer function of the system in air far from the sample,
we extract the resonance frequency $f_0^\infty$ and the quality factor $Q^\infty$.
The effective mass of the system is deduced from the knowledge of
the resonance frequency and the stiffness $k$, which is determined by
the thermal fluctuation method\cite{butt1995calculation}. 
The damping coefficient $\gamma$ is obtained through the quality factor
$Q^\infty$. The knowledge of these three quantities, $k$, $f_0^\infty$ and $Q^\infty$, is the only needed.

\section*{Results and Discussion}

\begin{figure}[ht]
\begin{centering}
\includegraphics[width=7.5cm]{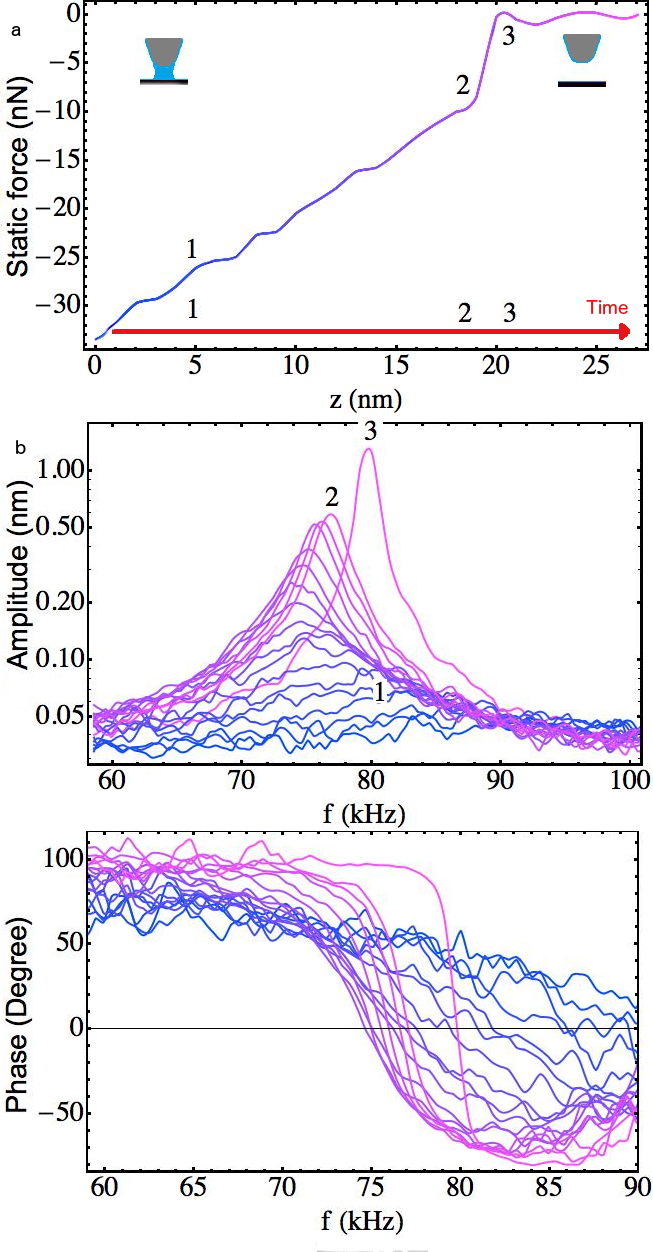}
\par\end{centering}
\caption{a) Static force of a capillary bridge measured by FFM during the retraction
of the sample. b)Viscoelastic properties of a water meniscus with
cantilever 1. System transfer functions are recorded during the retraction
of sample, from blue to pink. Transfer function after (3), before
(2) the capillary bridge breaks and close to the sample in presence
of the capillary bridge (1). The lever has a free resonance frequency
of 79 kHz and a quality factor in air of 127. The transfer function
clearly shows that the resonance frequency is shifting toward lower
frequency. This implies that the tip is probing an attractive interaction.
The resonance peak broadening and the flattening of the phase can
be interpreted as a strong dissipation.}
\label{fig:1}
\end{figure}

Fig~\ref{fig:1}b presents the amplitude and phase response when the tip is stopped
during retract in presence of a capillary bridge. We can associate
each spectrum to the static interaction, from blue to pink (see corresponding
numbering in the legend of Fig~\ref{fig:1}). The amplitude and phase
of the Fig.\ref{fig:1}b, which are the response of the system lever
plus capillary bridge at 79kHz, clearly show that the resonant frequency
is shifting towards lower frequencies meaning that an attractive force
gradient is exerted on the tip. The resonance peak broadening and
the flattening of the phase are representative of an increasing dissipation.

\begin{figure}[ht]
\begin{centering}
\includegraphics[width=8cm]{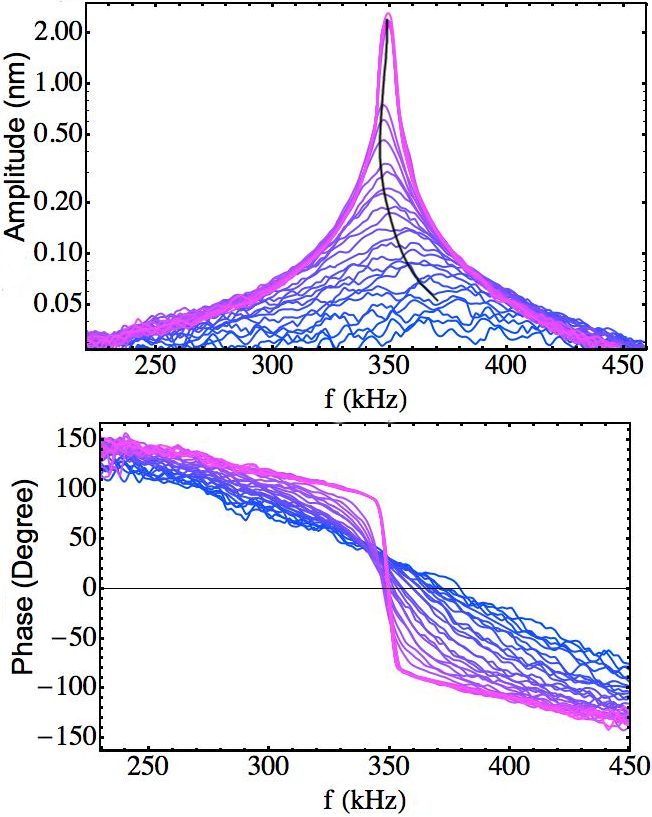}
\par\end{centering}
\caption{Viscoelastic properties of a water meniscus with cantilever 2, Spectra
are recorded during the retraction of sample, from blue to pink respectively
close to the sample and after the break of the capillary bridge. The
lever has a free resonance frequency of 349 kHz and a quality factor
in air of 200. The black line clearly shows that the resonance frequency
is shifting toward higher frequency. The resonance peak broadening
and the flattening of the phase can be interpreted as a strong dissipation.
\label{fig:2}}
\end{figure}

The results in Fig.\ref{fig:1}b contrast with those of the Fig~\ref{fig:2}
obtained using a higher frequency lever, with first eigenmode at 350kHz.
For this second lever, we observe that the resonance frequency is
shifting towards higher frequencies. This is the signature
of repulsive force gradient. 

\begin{figure}[h]
\begin{centering}
\includegraphics[width=8cm]{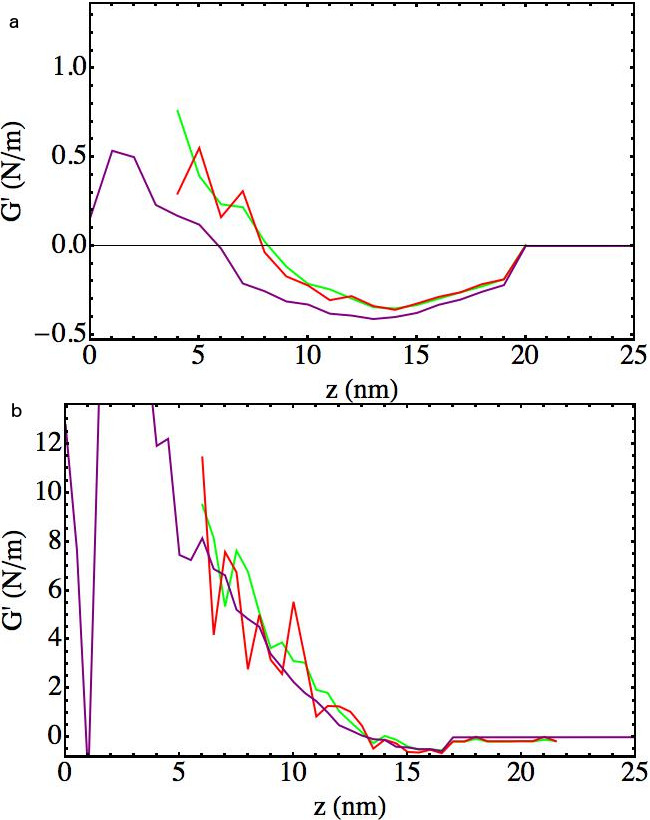}
\par\end{centering}
\caption{Stiffness of the interaction associated to the previous spectra (a)
for the 79 kHz lever and b)for the 350 kHz lever) evaluated in three
different ways. The green and red curves are obtained by fitting respectively
the amplitude response by Eq.\ref{eq:2 H(f)} and the phase response
by Eq.\ref{eq:1 phy}, to obtain the resonance frequency shift.Then
the stiffness of the interaction is calculated using Eq.\ref{eq:3}.
The purple curve is obtained using the linear transformation of the
amplitude and phase signal at the resonance frequency using Eq.\ref{eq:5}.
\label{fig:3}}
\end{figure}

\begin{figure}[h]
\begin{centering}
\includegraphics[width=8cm]{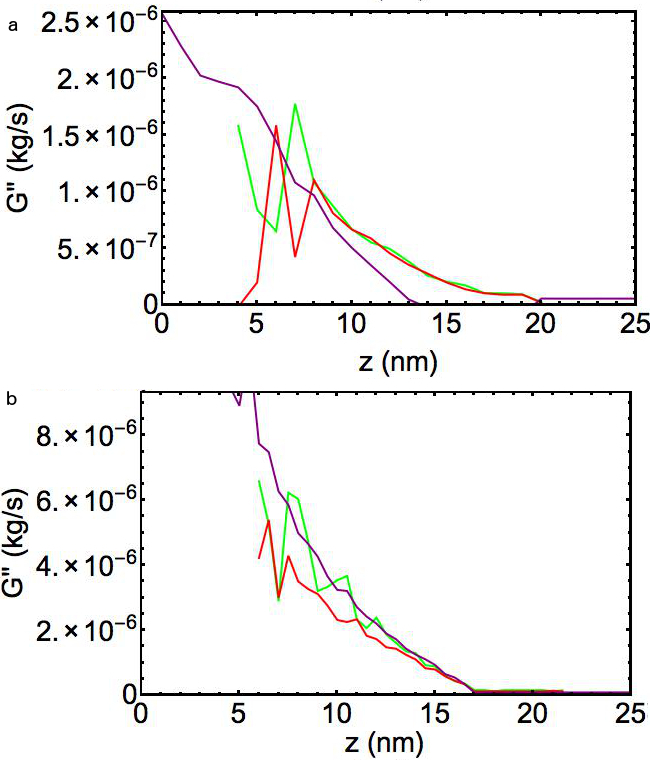}
\par\end{centering}
\caption{Dissipation $G'$ associated to the previous spectra (a)
for the 79 kHz lever and b)for the 350 kHz lever) evaluated in three
different ways. The green and red curves are obtained by fitting respectively
the amplitude response by Eq.\ref{eq:2 H(f)} and the phase response
by Eq.\ref{eq:1 phy}, to obtain the quality factor. Then the dissipation
of the interaction is calculated using Eq.\ref{eq:4}. The purple
curve is obtained using the linear transformation of the amplitude
and phase signal at the resonance frequency using Eq.\ref{eq:6}}
\label{fig:4}
\end{figure}

As explained in methods, the interaction stiffness $G'$ and the dissipation
$G''$ have been determined three times, one from
the amplitude curve (green Eqs \ref{eq:2 H(f)}, \ref{eq:3} and \ref{eq:4}), 
a second time from the phase curve (red Eqs \ref{eq:1 phy}, \ref{eq:3} and \ref{eq:4})
and a third time from the amplitude and phase changes at a fixed frequency
(purple, Eqs \ref{eq:5} trough \ref{eq:phy infiny}), presented in Figs~\ref{fig:3} and~\ref{fig:4}.
A first observation, is that the three methods give the same results. 

Fig.\ref{fig:3}a shows that the stiffness is negative about -0.5
N/m when the frequency is about 79kHz. Fig.\ref{fig:3}b presents
the stiffness measured with the second lever at 350kHz using the three
methods. It shows a strong repulsive interaction, up to 10N/m. The
dissipation for 350kHz and 79kHz is presented respectively in Fig.\ref{fig:4}a
and b. It exhibits similar values for the two frequencies.

In contrast with the measurement in our previous paper using a single lever, here
the experimental conditions cannot be the same when using different
levers. 
We however show in this paper that the two viscoelastic regimes found in our previous
article~\cite{carpentier2015out}, where another method is used, are corroborated using detection methods based on resonance
shift detection as classically done in AFM.


For low frequency, the nanomeniscus is attractive whereas at high frequency,
the nanomeniscus force gradient is repulsive. 
Direct contact between the tip and the sample can not
be an explanation of this strong repulsive stiffness. As we simultaneously
measure the static force and the dynamic response of the interaction
(see Fig.\ref{fig:1}a), we precisely know for each spectrum the tip-sample
distance and the static force acting on the tip. This strong effect
is only due to the capillary bridge behavior. 
At lower frequencies the bridge remains closer to equilibrium at constant 
pressure whereas at higher frequencies the contact line is
locked and the volume of the meniscus remains constant instead \cite{valsamis2013vertical}.

Fig.\ref{fig:4}b, which presents the dissipation when the system
is submitted to a repulsive interaction at 350kHz, enforces this conclusion
by showing a similar dissipation than the one in our previous article,
about 10\textsuperscript{-6}kg/s. In Fig.\ref{fig:3}a for the lever
with the first eigenmode frequency at 79kHz, the interaction stiffness
is again found negative using the three methods. However, the measured
stiffness is higher than the calculated gradient from the static force,
about -1N/m. This could be explained by the distorted z-axis due to
the thermal drift or by the fact that the molecule exchange process
has already decreased. Fig.\ref{fig:4}a presents the dissipation for
the lever at 79kHz. It is about 10\textsuperscript{-6}kg/s, which
corresponds to the same order of magnitude for the dissipation presented
in our previous paper when the contact line is partially locked at
high frequency \cite{carpentier2015out} and in Fig.\ref{fig:4}b.
Using results from our previous paper it appears to us that the nanobridge
behavior at 79kHz is no longer in thermodynamical equilibrium. 

\section*{Conclusion}

In conclusion, using classical AFM modes, i.e. frequency shift and
amplitude and phase changes at constant frequency of excitation, we have
again found a transition in the visco-elastic properties of a water nanobridge
between thermodynamical equilibrium and out of equilibrium from attractive to repulsive behavior respectively. 
In this out of equilibrium regime, we interpret the present results using a constant volume nanobridge
(no molecular exchange between vapor and nanobridge) and a fixed contact
line \cite{carpentier2015out}. 

This study has been made possible using low spring constant cantilever and the FFM scheme: a feedback
force applied to the tip in real time avoids the ``jump to contact'' due to the nucleation of the bridge.
The results obtained are fully consistent with the ones presented before. 
The key advantage of the method used in our previous paper is that it allows to work at any
frequency which enables us to use a single lever when changing the
excitation frequency from low to high frequencies. Here, working only
at lever resonance and in close proximity to the resonance frequency, we have
not been able to reach thermodynamical equilibrium as shown
in the previous paper.

\section*{Acknowledgments}
Mario S. Rodrigues and Miguel V. Vitorino acknowledge financial support
from Fundação para a Ciência e Tecnologia, grants SFRH/BPD/69201/2010
and PD/BD/105975/2014 respectively. We would like to acknowledge Jean-Francois
Motte who gave precious help and advices on this experiment.

%
%
%

\end{document}